\documentstyle[epsf,aps,prl,multicol]{revtex}
\begin{document}
\draft
\title{
Fluctuations in quantum dot charging energy}
\author{M. Stopa} 
\address{RIKEN (The Institute of Physical
and Chemical Research)
2-1, Hirosawa, Wako-Shi
Saitama 351-01, Japan \\
e-mail stopa@sisyphus.riken.go.jp}
\date{\today}
\maketitle

\begin{multicols}{2}
[
\begin{abstract}
We demonstrate that enhanced fluctuations in the charging energy
of chaotic semiconductor quantum dots result from strongly scarred
wave functions. We demonstrate that strong scars linger at the Fermi
surface due to their enhanced Coulomb interaction between spin
up and spin down states.
We present density functional and spin density 
functional calculations to elucidate the temperature, dot shape
and spin polarization dependencies of the fluctuations.
\end{abstract}
]


A simple, physically appealing paradigm for the statistical
behaviour of the ground state of chaotic, 
interacting systems does not yet exist.
Based on the concept of level repulsion, the behaviour of
eigenvalues and vectors of complex systems
has been successfully treated 
for many years via random matrix theory (RMT) \cite{Brody}.
As recent experiments on quantum dots demonstrate however \cite{Sivan},
fluctuations in the ground state energy 
as a function of electron number $N$, are not well 
described by RMT since the required hypothesis of
constant interaction (CI) appears to be violated.
On fundamental grounds this failure is to be expected
since the ergodicity implicit in RMT is unconstrained by
any minimization principle which must be an essential
ingredient of any theory of the ground state.

The charging energy $E_C$ of a quantum dot, defined below in terms of
the ground state energy at differing $N$, provides an important
archetype of an interacting system insofar as particle
number, disorder \cite{LP1},
the ``chaoticity'' \cite{Marcus} and size can be varied.
Attempts to understand the scale and behaviour of fluctuations in
the charging energy in quantum dots have included
numerical studies of tight binding Hamiltonians for small particle 
number \cite{Sivan,Berkovits} as well as study of the statistical
behaviour of electron-electron interactions in a 
circular, disordered dot based on the random phase 
approximation (RPA) for the screened interaction and a sigma
model calculation for the 
eigenfunction correlator \cite{Blanter}.

Exact diagonalization studies such as in Refs. \cite{Sivan,Berkovits}
give valuable insight into trends in the ground state energy, 
but they
are necessarily limited to small particle number and in addition they 
depend on a reduction of a complicated set of matrix
elements to a small set of energy parameters. Reference \cite{Blanter},
on the other hand, considers explicitly direct and exchange Coulomb
matrix elements, however the distribution of these elements are
evaluated using a wave function correlator for a non-interacting
disordered system. Consequently there is no minimization principle
and no self-consistency in the theory. Furthermore, both the
diagonalization approach and the method of Ref. \cite{Blanter}
provide only statistical information on the charging energy.
Neither method can realistically incorporate a varying gate voltage 
and therefore the physical {\it process} whereby
fluctuations occur is unapproachable. 

\newcommand{\Fpp}{$\partial^2 F(N,V_g)/\partial N^2$}
\newcommand{\bfr}{{\bf r}}

By contrast, a detailed study of the evolving, self-consistent
eigenfunctions, free energy and level structure of a realistic
dot via density functional (DF) theory, 
which we present here, reveals the specific mechanism
whereby $E_C$ fluctuations deviate from the ``RMT+CI''

 \begin{figure}[hbt]
 \setlength{\parindent}{0.0in}
  \begin{minipage}{\linewidth}
 \epsfxsize=8cm
 \epsfbox{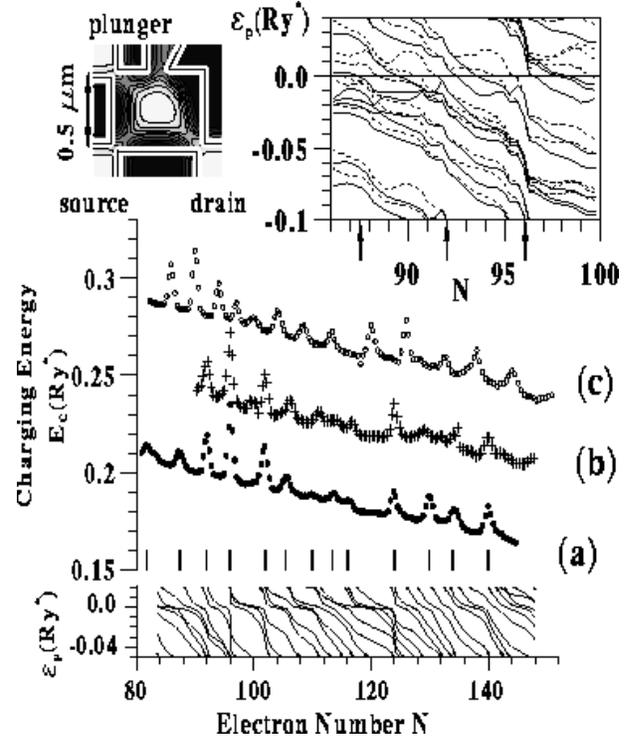}
 \vspace*{3mm}
 \caption{$E_C$ vs. $N$ and spectrum
 near $E_F$ (lower two panels). Curves (a) and (b) differ from (c)
 by non-plunger gate voltages, hence varied pattern of peaks. (b)
 and (c) offset by $0.04$ and $0.08 \; Ry^*$ resp.
 (a) and (c) are pure Hartree, (b) includes $E_{xc}$
 though LDA. $T=0.1 \; K$ throughout. 
 Spectrum corresponds to (a) and (b). Upper right panel shows levels 
 from SDF calculation,
 same parameters as (b), arrows indicate $E_C$ peaks. Note at $N=92$ and $96$
 spin polarization of dot collapses. Inset: gate pattern and typical
 self-consistent effective 2D potential at 2DEG level.}
  \end{minipage}
 \end{figure}

\noindent predictions. We show that large fluctuations in $E_C$
are related to strong scars; remnants of periodic orbits
in the classical confining potential. When located at the Fermi
surface $E_F$ these states, being quasi-one dimensional, tend to
create an inhomogeneous potential, in contrast to other
more ``chaotic'' states which occupy more smoothly
the entire dot area. Consequently, when a scar is at
the Fermi surface a gap is created to the next state
which results in additional gate voltage spacing to the next
Coulomb oscillation. Furthermore, we find that the direct Coulomb
matrix element between a state and itself (up spin and down spin),
denoted $W_{pp} \equiv U_p$, is generally greater than between
different spatial states $W_{pq}, \; p \ne q$, 
and that $U_p$ for scars is particularly large. Therefore,
as we verify via full spin density functional (SDF) calculations
for certain parameter regions, the dot is almost always
spin polarized \cite{Marcus2}.
Further, scars tend to remain, half filled, 
at $E_F$ (as $V_g$ is varied)
due to their large $U_p$, while other, more homogeneous states
pass through the Fermi surface \cite{correlation} 
(in the process gradually screening
the scar state). Finally, when it becomes energetically 
impossible to draw down a still higher state to $E_F$, the filling
of the second spin state of the scar does occur, leaving a gap
to the next state, as mentioned above. At this point not only
the scar, but all states are doubly occupied and spin polarization of
the dot collapses.

We calculate the self-consistent electronic structure
\cite{LP1} for a small quantum dot with the wafer
profile and gate pattern of the device used by Sivan {\it et al.}
\cite{Sivan}. The nominal two dimensional electron gas
(2DEG) density is 
$3.1 \times 10^{-11} cm^{-2}$. Typical values for 
the average level spacing (not assuming spin degeneracy) 
and charging energy are $\Delta \sim 0.012 \; Ry^*$ and
$E_C \sim 0.18 \; Ry^*$, respectively ($1 \; Ry^* \approx 5.8 \; meV$).
The gate pattern is
rectangular (inset Fig. 1) but due to excess metal
in one gate the effective confining potential 
is classically chaotic, a fact which is substantiated
by the statistics of the single particle level spacings
(not shown). As in the experiment we vary the ``plunger'' gate
voltage $V_g$ (inset Fig. 1), typically in steps of $5 \; mV$,
the other gates being fixed 
to isolate the dot from source and drain.
For each $V_g$ the electronic structure is calculated at three
values of 
dot chemical potential $\mu_{dot}= 0.0,\pm 0.2 \; Ry^*$. 
States are filled (and $N$ thereby determined)
according to a Fermi function, which except
for very small dots is expected to give a good approximation
to the full grand canonical ensemble \cite{Jovanovic}. Thus, in
the calculation $N$ is not necessarily an integer. The values
of $\mu_{dot}$ are chosen to change $N$ by about $\pm 1$, relative
to the ``equilibrium'' value ($\mu_{dot} = \mu_{leads} \equiv 0$).
The free energy $F(N,V_g)$ is
calculated according to ref. \cite{LP1} and $E_C$ is
{\it defined} as \Fpp, which we
compute discretely.

Traces of $E_C (N)$ ($N$ from $\mu_{dot}=0$ solution)
are shown in Fig. 1. The quasi-periodic peaks are associated
with the onset (e.g. $N = 87, \; 114$) 
or completion ($N=92, \; 96, \; 102, \;
124$) of scar filling, the former being accompanied
by downward fluctuation of states below the scar state, the latter with gaps
above the scar state.
Changing ``non-plunger'' gates varies the precise pattern of
oscillations, but not the basic structure. Varying a ``back gate'' (not shown)
also produces qualitatively similar characteristics. 

 \begin{figure}[hbt]
 \setlength{\parindent}{0.0in}
  \begin{minipage}{\linewidth}
 \epsfxsize=8cm
 \epsfbox{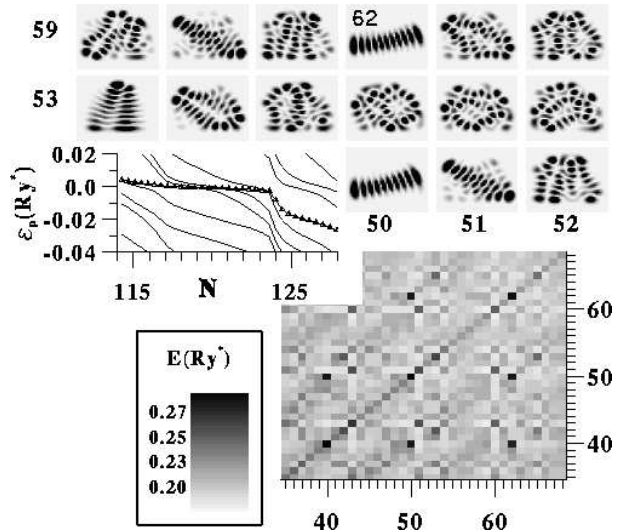}
 \vspace*{3mm}
 \caption{Selected eigenfunctions (moduli squared) and grey
 scale of Coulomb matrix elements at $N \approx 124$ (cf. curves (a)
 and (b) Fig. 1). Inset gives close-up of spectrum from Fig. 1, scar
 state indicated by triangles remains near $E_F$.}
  \end{minipage}
 \end{figure}
 
Standard DF theory is a spin degenerate mean field
theory. Even when exchange-correlation $E_{xc}$ is included in the
local density approximation (LDA) it is known that some uncorrected 
self-interaction remains \cite{Slater_Zunger}. 
The tendency (Fig. 1) for levels to cluster at $E_F$ results from
the energetic advantage of occupying several states partially
rather than one or two states fully. At finite temperature $T$
such partial occupancy can be justified as an approximation to 
the weighted ensemble of many-body states. Further, in this
regime, Wigner crystallization is unimportant. 
Nonetheless, a more precise description, where the
role of spin polarization is made manifest, can be obtained via
SDF calculations. As these are numerically more intensive we
only calculate the behaviour for various representative parameter
regions. Our main physical conclusion is that the electrostatically
driven tendency for levels to cluster, partially filled, at $E_F$
in ordinary DF theory is replaced by an electrostatically driven
spin polarization in SDF (e.g. it is energetically favorable to occupy
two different spatial states singly as opposed to occupying
either of them doubly).

The spectrum for the SDF case
is illustrated in the upper right panel of Fig. 1. 
The $E_C$ fluctuations (not shown) occur at the same $N$
values as in the DF calculations and coincide
with the arrows. Note that the fluctuation at $N=87$, which corresponds
to the filling of the first spin state of a scar, is qualitatively
different from those at $N=92$ and $96$, where scars become
doubly occupied. In the latter cases, since $U_p$ of the scar is
much greater than that of other states, double filling
generally only occurs when all other states are either doubly
filled or empty. Therefore
\cite{Pablo}, when this state fills, all states of the dot
become identically spin degenerate. 

Figure 2 shows wave functions and Coulomb
matrix elements in the vicinity of a typical $E_C$ fluctuation,
$N \approx 124$. 
Coulomb matrix elements are calculated using the kernel of
Poisson's equation from the self-consistent calculation. Thus screening
by the gates is included automatically \cite{future}.
We have identified four stable orbits for this device
of which two are observable in Fig. 2. Here $n_{62}=2$ ($n_p \equiv $ occupancy
of state $p$) and $n_{63}=0$. 
The grey scale plot of Coulomb matrix elements 
shows that this state, as well as $p=60$, another scar along a slightly 
different orbit, 
have higher values of $U_p$ \cite{Bfield}. 
Scars occur
in families as additional nodes are added along the same orbit
(compare $p=62$ and $p=50$ and note their interaction). The 
Coulomb matrix elements given here  
\begin{figure}[hbt]
 \setlength{\parindent}{0.0in}
  \begin{minipage}{\linewidth}
 \epsfxsize=8cm
 \epsfbox{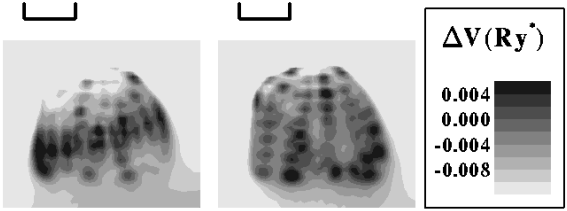}
 \vspace*{3mm}
 \caption{Potential difference contours showing influence
 of $p=62$, scar state filling (left) and that of
 more chaotic state $p=64$ (cf. Fig. 2). Plunger gates indicated.}
  \end{minipage}
 \end{figure}
\noindent are between self-consistent
states  \cite{note1}.
Thus, even though
$W_{62,63}$ is relatively small ($\sim 0.18 \; Ry^*$),
it is the interaction of these two states which produces the gap
at $E_F$ and the fluctuation of $E_C$ at this $N$. The 
strong 1-d scar $p=62$ at $E_F$ acts like a single atomic valence electron and
strongly influences the potential (Fig. 3),
creating a central ``ridge'' which modifies $p=63$, reducing the Coulomb
interaction between $p=62$ and $p=63$.
 
A comparison between our calculations and the experimental 
results of Sivan {\it et al.} \cite{Sivan} shows a striking discrepancy. 
While the gate voltage spacing fluctuations ($\Delta_2$ in Ref. \cite{Sivan})
are symmetric, the fluctuations of $E_C$ 
computed here all proceed upward from a smoothly decreasing base.
Furthermore, a comparison of the histograms of the fluctuations
(not shown) reveals that the fluctuations in $E_C$ have an
r.m.s. deviation of only about $5 \% $ of $E_C$, a result that is
smaller by a factor of two or three from the data
for $\Delta_2$. These discrepancies emphasize the importance,
heretofore not fully appreciated, of distinguishing between the
``inverse compressibility'' (what we call the charging energy)
at fixed $V_g$ and the gate voltage spacing (normalized by the 
capacitance ratio $\alpha$, see below) between Coulomb oscillations.
These two quantities are equivalent only within the constant interaction
model and, as we now show, small, upward variations in $E_C$ can produce
much larger, symmetric variations in $\Delta V_g$. 

To a reasonable approximation we can write 
$F(N,V_g)=\frac{1}{2} E_C(V_g) N^2 + b(V_g)N +c$,
i.e. $F$ is quadratic in $N$ at each $V_g$. In the CI model, 
$b(V_g)=-e \alpha V_g$ where $\alpha$ is the ratio of the dot-gate
capacitance $C_{dg}$ to the dot self-capacitance $C$, and $E_C = \frac{e^2}{C}
+ \sum_p^N \epsilon^0_p$ (at $T=0$). $C$, $\alpha$ and $\epsilon^0_p$ 
(the single particle level energies) are all
assumed to be independent of $V_g$. $\Delta V_g$ is determined by writing
$F(N,V_g)=F(N+1,V_g)$ and $F(N+1,V_g^{\prime})=F(N+2,V_g^{\prime})$ and
subtracting, resulting in: 
\begin{equation}
e \alpha \Delta V_g = \frac{e^2}{C} + 
\epsilon^0_{N+2}-\epsilon^0_{N+1} \label{eq:std}.
\end{equation} 
However when $C$, $\alpha$ and $\epsilon^0_p$ depend on
$V_g$, the procedure for deriving Eq. \ref{eq:std}
leads to
\begin{eqnarray}
e \alpha(V_g^{\prime}) V_g^{\prime} - e \alpha(V_g) V_g  & = &
\frac{(N+3/2)e^2}{C(V_g^{\prime})}  \nonumber \\ 
- \frac{(N+1/2)e^2}{C(V_g)} &  &
+\epsilon^0_{N+2}(V_g^{\prime})  - \epsilon^0_{N+1} (V_g). \label{eq:compl}
\end{eqnarray}
When, in particular, $C$ fluctuates appreciably from $V_g$ to $V_g^{\prime}$
the influence on $\Delta V_g$ can be magnified, as illustrated in
Fig. 4. Here we have calculated $F(N,V_g)$ on sufficiently fine
grids to determine $\Delta V_g$ numerically.

To complete the picture we note that the frequency of scars
in the spectrum, and hence the number of large oscillations
in $E_C$, decrease with $N$. This is somewhat evident in Fig. 1 
in that the spacing of oscillations increases with $N$. 
The scars are remnants of closed orbits and are thus effectively
one dimensional. Their density of states (DOS) decreases as $1/\sqrt{E}$,
in contrast to the constant DOS for the 2D dot as a whole. Thus
the overall frequency of
scars in the spectrum will diminish. 
In addition, the extent to which a scar at $E_F$ pushes up the 
ensuing state should also diminish with $N$. (Again, the oscillations in
Fig. 1 for larger $N$ appear to grow smaller). The reason is twofold. First,
screening by the dot electrons increases with $N$. Second, the influence of 
the inhomogeneous potential fluctuation produced by the scar 
on the energy of a typical chaotic state should depend on the
wavelength of that state $\lambda$ in comparison to the dot size, i.e. the kinetic
energy cost of avoiding the scar decreases with $\lambda$.

When one considers larger dots with complicated scars occupying 
much of the dot area, the distinction between scars and non-scars,
in terms of Coulomb matrix elements, tends to disappear
\cite{note2}. In this case the 
fluctuations in $E_C$ will depend on the dot size, the fraction
of the dot area occupied by the states near $E_F$, the screening length
$\lambda_{sc}$ and perhaps dot geometry.

Sivan {\it et al.} found that while the fluctuations in 
$e \alpha \Delta V_g$ were much greater than the average single particle level
spacing $\Delta$, the temperature dependence of these fluctuations was,
seemingly, governed by $\Delta/k_B T$ \cite{Sivan}. We have argued that the 
fluctuations in $E_C$ should decrease with $N$ due to decreasing frequency 
of scars and their 

 \begin{figure}[hbt]
 \setlength{\parindent}{0.0in}
  \begin{minipage}{\linewidth}
 \epsfxsize=8cm
 \epsfbox{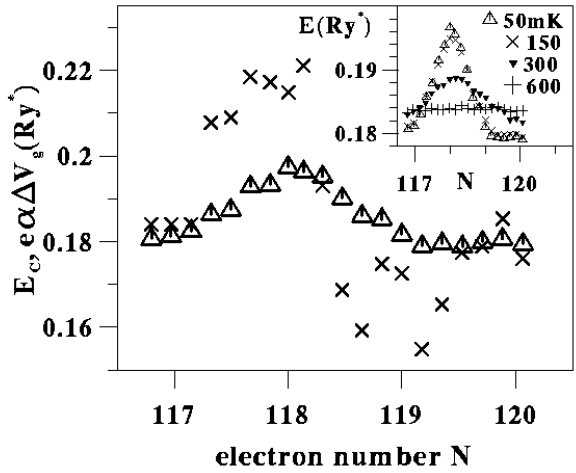}
 \vspace*{3mm}
 \caption{Charging energy and gate voltage spacing vs $N$.
 Non-plunger gates differ from preceding plots, giving large fluctuation
 at $N=118$.
 While $E_C$ fluctuation is purely upward, fluctuation in 
 $V_g$ spacing is symmetric and substantially greater in
 magnitude. Inset: $E_C$ fluctuation for various $T$. Fluctuations
 vanish far below $k_B T = E_C$.}
  \end{minipage}
 \end{figure}

\noindent decreasing influence on neighboring states. We cannot say
definitively, however, whether these fluctuations should scale with
$\Delta$ (dot area) or with $E_C$ (dot radius), though since
$\lambda_{sc} \sim \Delta^{-1}$, the former
seems more likely in the large dot limit.
Empirically, however, the fluctuations in $E_C$ are comparable to 
$\Delta$ and, as noted, the enhancement of fluctuations in $\Delta V_g$
(Eq. \ref{eq:compl}) derives from $V_g$ dependence of capacitances
and energy levels.
Consequently, (Fig. 4 inset),
the fluctuations essentially vanish by $600 \; mK$ ($\approx 0.75 \Delta$)
even though this is a factor of twenty smaller than $E_C$ itself.

Finally, in recent studies by Patel {\it et al.},
citation in \cite{Marcus2}, no correlation was found to exist between peak 
heights and
spacings. We propose that a stadium geometry, with quantum point contacts
(QPCs) placed along the long axis, should indeed show this correlation. Certainly
one stable periodic orbit will be that which simply oscillates along the
central axis, so that these states will give greatly enhanced 
peak heights. Since, according to our discussion,
scars associated with this orbit will float at $E_F$ for certain ranges of
$V_g$, one should find corresponding groups of large Coulomb oscillations
separated by a large $V_g$ spacing
from neighboring oscillations.

In conclusion, we have demonstrated that strongly scarred, quasi-1d states,
by virtue of their strong Coulomb interactions, produce quasi-periodic 
fluctuations in the charging energy of chaotic quantum dots. We have shown
that the scale of those fluctuations is consistent with the single particle 
level spacing, but that fluctuations in gate voltage spacings of Coulomb
oscillations are, due to $V_g$ dependence of capacitances and energy levels,
substantially greater and differently (symmetrically) distributed. Finally,
we have proposed a simple experimental arrangement which could 
elucidate the role of self-consistency in chaotic systems discussed herein.

I wish to thank Dr. Uri Sivan for providing me with details
of his devices and measurement. I wish to acknowledge 
a particularly helpful conversation with Oded Agam. I have also had
helpful conversations with
Y. Alhassid and co-workers, S. Das Sarma, F. Stern,
C. Marcus, S. Patel, N. Wingreen
Y. Aoyagi and T. Sugano. Support from Riken
Computing Center - Fujitsu VPP500 Supercomputer 
are gratefully acknowledged. This work supported by grants from the
Japan Ministries of Education and Science and Technology.

\end{multicols}


\begin{references}

\bibitem{Brody} T. A. Brody {\it et al.} Rev. Mod. Phys. {\bf 53}, 385 (1981);
F. Haake, {\it Quantum Signatures of Chaos} (Springer,
Berlin, 1991).

\bibitem{Sivan} U. Sivan {\it et al.} Phys. Rev. Lett. 
{\bf 77}, 1123 (1996).

\bibitem{LP1} M. Stopa, Phys. Rev. B, {\bf 54}, 13767 (1996).

\bibitem{Marcus} J. A. Folk {it et al.}
Phys. Rev. Lett. {\bf 76}, 1699 (1996).

\bibitem{Berkovits} R. Berkovits and Y. Avishai, J. Phys. Condens.
Matter {\bf 8} 389 (1996).

\bibitem{Blanter} Ya.M. Blanter {\it et al.}
Phys. Rev. Lett. {\bf 78}, 2449 (1997).

\bibitem{Marcus2} This explains the absence of an even-odd
symmetry which would be expected in a spin degenerate
version of the CI model, 
S. R. Patel {\it et al.}, preprint, cond-mat/9708090.

\bibitem{correlation} The self-consistent tendency for multiple levels
to be partially occupied near $E_F$ for non-zero $T$ also
provides a physical explanation for unexpected correlation of
peak heights; see M. Stopa, Phys. Rev. B {\bf 48}, 18340 (1993)
and, more recently, J. A. Folk {\it et al.}, Phys. Rev. Lett. {\bf 76},
1699 (1996).

\bibitem{Jovanovic} D. Jovanovic and J. P. Leburton, Phys. Rev B,
{\bf 49}, 7474 (1994).

\bibitem{Slater_Zunger} J. C. Slater, 
{\it The Self-Consistent Field for Molecules
and Solids}, McGraw-Hill, New York, 1974. More recent techniques on
elimination of self-interaction effects in DF theory are discussed in
M. M. Rieger and P. Vogl, Phys. Rev. B {\bf 52} 16567 (1995) and
references therein.

\bibitem{Pablo} P. I. Tamborenea, R. J. Radtke and S. Das Sarma,
preprint, cond-mat/9604099.

\bibitem{future} M. Stopa, (to be published).

\bibitem{Bfield} We note that the application of a magnetic field $B$, 
by reducing the effective dimensionality of all eigenstates also 
imparts structure to the Coulomb matrix elements. The influence of
$B$ on the spectrum can be seen in Ref. \cite{LP1}, Fig. 13.

\bibitem{note1} This is to be distinguished from the Coulomb matrix 
elements, between bare levels, screened by the dot electrons themselves,
as determined, for example, via RPA. 

\bibitem{note2} For larger $N$ one expects scars to be associated
with groups of eigenstates rather than our small dot case 
where they are represented by single
eigenfunctions.

\end{references}
\end{document}